\documentclass[aps,prb,preprint,groupedaddress]{revtex4}

\usepackage{graphicx}
\usepackage{dcolumn}
\usepackage{bm}
\usepackage{amsmath,amssymb}

\def\cH{{\mathcal H}}
\def\cHeff{{\mathcal H}_{\rm eff}}
\def\vS{\mbox{\boldmath $S$}}
\def\vT{\mbox{\boldmath $T$}}
\def\JF{J_{\rm F}}
\def\JAF{J_{\rm AF}}
\def\DAF{\Delta_{\rm AF}}
\def\DF{\Delta_{\rm F}}

\def\dstyle{\displaystyle}

\begin{document}


\title{Ground-State Phase Diagram of an Anisotropic \mbox{\boldmath $S=1$} Ferromagnetic-Antiferromagnetic
Bond-Alternating Chain}



\author{Kiyomi \textsc{Okamoto}$^{1}$, Takashi \textsc{Tonegawa}$^{2,3}$, Makoto \textsc{Kaburagi}$^{2}$
 and T\^oru \textsc{Sakai}$^{1,4}$}
\affiliation{
$^{1}$Graduate School of Material Science, University of Hyogo, Hyogo 678-1297, Japan \\
$^{2}$Professor Emeritus, Kobe University, Kobe 657-8501, Japan \\
$^{3}$Department of Physical Science, Osaka Prefecture University, Osaka 599-8531, Japan \\
$^{4}$National Institutes for Quantum and Radiological Science and Technology (QST), SPring-8, Hyogo 679-5148, Japan
}


\date{August 28, 2019}

\begin{abstract}

By using mainly numerical methods, 
we investigate the ground-state phase diagram (GSPD) of an $S=1$
ferromagnetic-antiferromagnetic bond-alternating chain with the $XXZ$ and the on-site anisotropies.
This system can be mapped onto an anisotropic
spin-2 chain when the ferromagnetic interaction is much stronger than the antiferromagnetic interaction.
Since there are many quantum parameters in this system,
we numerically obtained the GSPD on the plane of the magnitude of the antiferromagnetic coupling
versus its $XXZ$ anisotropy,
by use of the exact diagonalization, 
the level spectroscopy as well as the phenomenological
renormalization group.
The obtained GSPD consists of six phases.
They are the $XY$1, the large-$D$ (LD), the intermediate-$D$ (ID), 
the Haldane (H), the spin-1 singlet dimer (SD), and the N\'eel phases.  
Among them, the LD, the H, and the SD phases are
the trivial phases, while the ID phase is the symmetry-protected topological phase. 
The former three are smoothly connected without any quantum phase transitions.
It is also emphasized that the ID phase
appears in a wider region compared with the case of the GSPD
of the anisotropic spin-2 chain with the $XXZ$ and the on-site anisotropies.
We also compare the obtained GSPD with the result of the perturbation theory.
\end{abstract}


\maketitle


\section{Introduction}

In recent years, low dimensional quantum spin systems have been
attracting increasing attention because they provide rich
physics even when models are rather simple.
Several years ago, we investigated\cite{tone,oka1,oka2,oka3} the $T=2$ quantum spin chain
 with the $XXZ$ and on-site anisotropies
described by
\begin{equation}
    \cH_1
    = \sum_j (T_j^x T_{j+1}^x + T_j^y T_{j+1}^y +\Delta T_j^z T_{j+1}^z)
      + D_2 \sum_j (T_j^z)^2,
    \label{eq:Ham-1}
\end{equation}
where $T_j^\mu$ ($\mu = x,y,z$) represents the $\mu$-component of the spin-2
operator $\vT_j$ at the $j$th site, and $\Delta$ and $D_2$ are, respectively, the
$XXZ$ anisotropy parameter of the nearest-neighbor interactions and the on-site anisotropy parameter.
We obtained the ground-state phase diagram\cite{tone,oka1} (GSPD) mainly by the use of the exact diagonalization and the
level spectroscopy (LS) analysis\cite{okamoto-ls,nomura-ls,kitazawa-ls,nomura-kitazawa-ls}.
The remarkable features of the GSPD are: 
(a) there exists the intermediate-$D$ (ID) phase which was first
predicted by Oshikawa\cite{oshikawa} in 1992 and has been believed to be
absent for about two decades until our finding\cite{tone,oka1,oka2} in 2011;
(b) the Haldane (H) state and the large-$D$ (LD) state belong to the same phase.
These features are consistent with the discussion by Pollmann {\it et al.}\cite{pollmann2010,pollmann2012}.
Namely, the ID state is a symmetry-protected topological (SPT) state protected by 
(i) the time-reversal symmetry $\vS_j \to -\vS_j$, as well as by
(ii) the space inversion symmetry with respect to a bond,
while the H state and the LD state are trivial states.
Slightly after our works,
the ID phase was also discussed by Tzeng \cite{tzeng} and Kj\"all et al. \cite{kjall}.

Considering these situations, 
we investigate the GSPD of the $S=1$ ferromagnetic-antiferromagnetic bond-alternating chain,
since it is thought that this chain can be mapped onto the spin-2 model
in the strong ferromagnetic coupling limit.
We describe our model in \S2, and the numerically determined GSPD are shown in \S3.
In \S4 the perturbation theory from the strong ferromagnetic coupling limit is developed.
Section 5 is devoted to concluding remarks.

\section{Model}

We investigate the model described by
\begin{eqnarray}
  &&{\cal H} = \sum_j \{ h^{\rm F}_{2j-1,2j}
                   + h^{\rm AF}_{2j,2j+1} 
                   + D_2(S_j^z)^2 \} \,, 
  \label{eq:our-ham}                 \\
  && h^{\rm F}_{j,j'} = -J_{\rm F} (S_j^x S_{j'}^x + S_j^y S_{j'}^y 
                       + \Delta_{\rm F} S_j^z S_{j'}^z)\,,  \\
  && h^{\rm AF}_{j,j'} = J_{\rm AF} (S_j^x S_{j'}^x + S_j^y S_{j'}^y
                       + \Delta_{\rm AF} S_j^z S_{j'}^z)\,.
\end{eqnarray}
\noindent
Here, $S_j^\mu$ (\hbox{$\mu=x$}, $y$, $z$) is the $\mu$-component of the
spin-1 operator $\vS_j$ acting on the $j$th site; $J_{\rm F}\,(>0.0)$ and
$J_{\rm AF}\,(\geq 0.0)$ denote, respectively, the magnitudes of exchange
interaction constants for the ferromagnetic and antiferromagnetic bonds;
$\Delta_{\rm F}$ and $\Delta_{\rm AF}$ are, respectively, the parameters
representing the $XXZ$ anisotropies of the former and latter interactions.

\begin{figure}[ht]
   \begin{center}
      \includegraphics[scale=0.35]{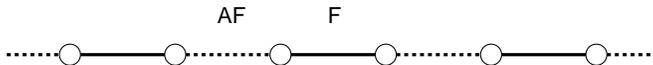}
   \end{center}
   \caption{A sketch of the present model.
            Open circles denote $S=1$ spins.
            Solid and dotted lines are the ferromagnetic (F) and antiferromagnetic (AF) bonds,
            respectively.}
   \label{fig:model}
\end{figure}

\section{Ground-State Phase Diagram by Numerical Calculations}

Since there are five parameters, $\JF$, $\JAF$,
$\DF$, $\DAF$, and $D_2$,
in our Hamiltonian (\ref{eq:our-ham}),
we restrict ourselves to the case
where $\JF=1.0$ (namely, $\JF$ is the unit of energy), 
$\DF=0.8$, and $D_2=-1/30$, 
and numerically determine the GSPD on the $\JAF$ versus $\DAF$ plane. 

\begin{figure}[ht]
   \begin{center}
      \includegraphics[scale=0.4]{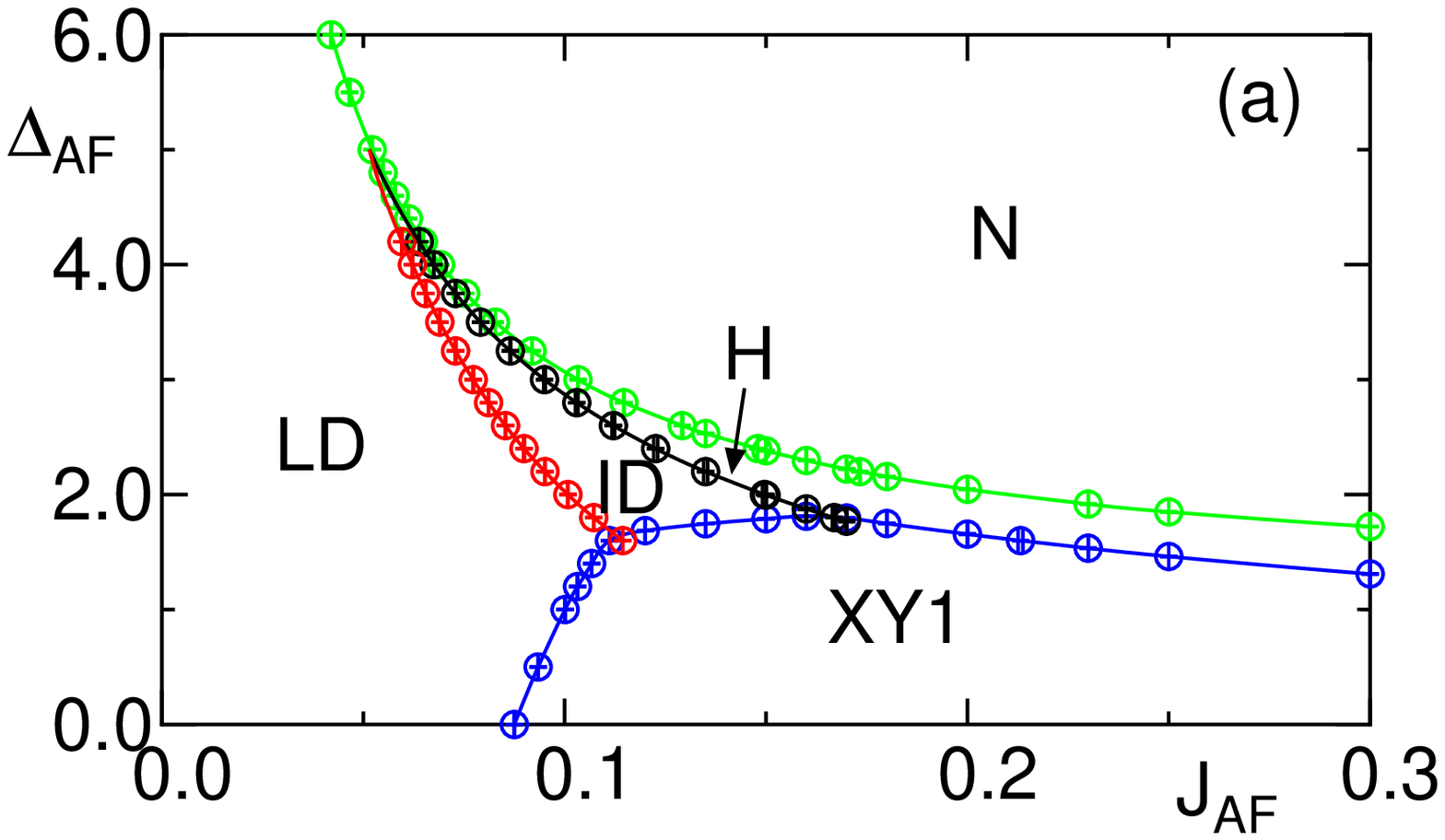}~~~~~
      \includegraphics[scale=0.4]{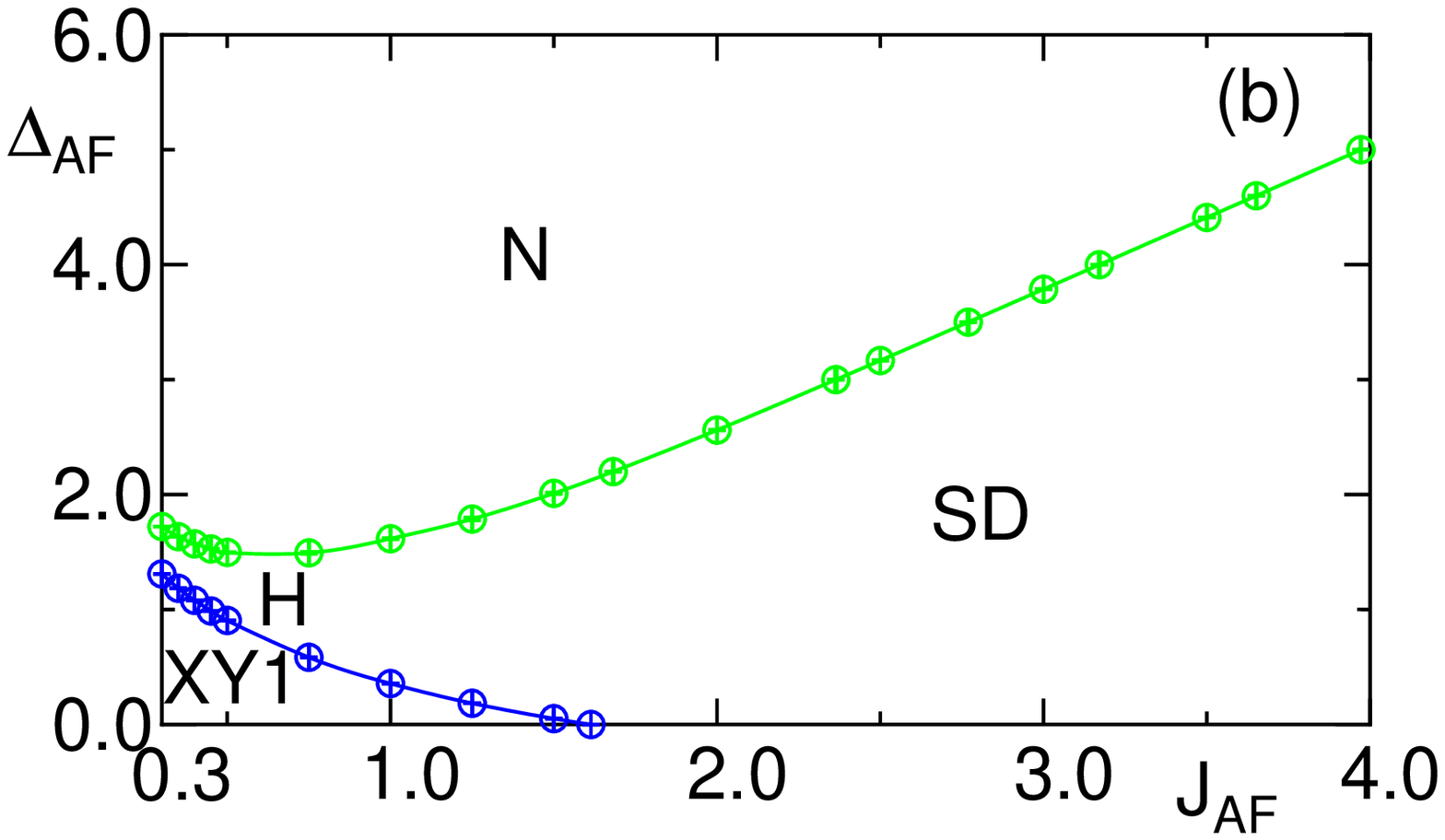}
   \end{center}
   \caption{The GSPD on the $\JAF$ versus
            $\DAF$ plane for the case of $\JF = 1.0$, $\DF = 0.8$, and $D_2 = -1/30$, 
            obtained in the present work;~(a) $0.0 \le \JAF \le 0.3$,
            (b) $0.3 \le \JAF \le 4.0$.}
   \label{fig:fig2}
\end{figure}

\begin{figure}[ht]
   \begin{center}
      \raisebox{0.2cm}{\texttt (a)} \includegraphics[scale=0.25]{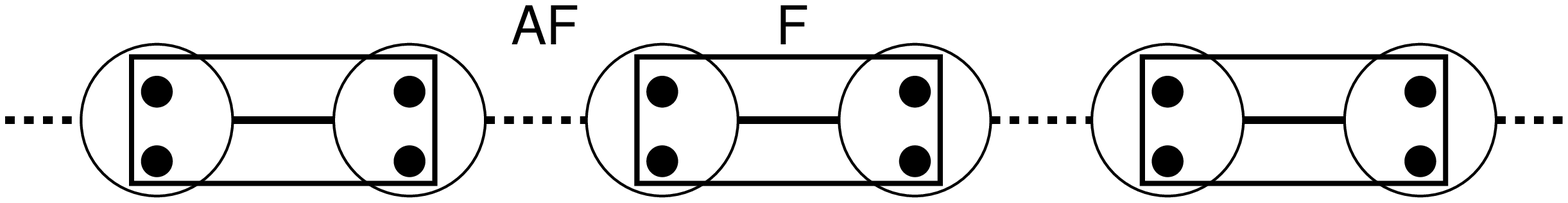}~~~~~
      \raisebox{0.2cm}{\texttt (b)} \includegraphics[scale=0.25]{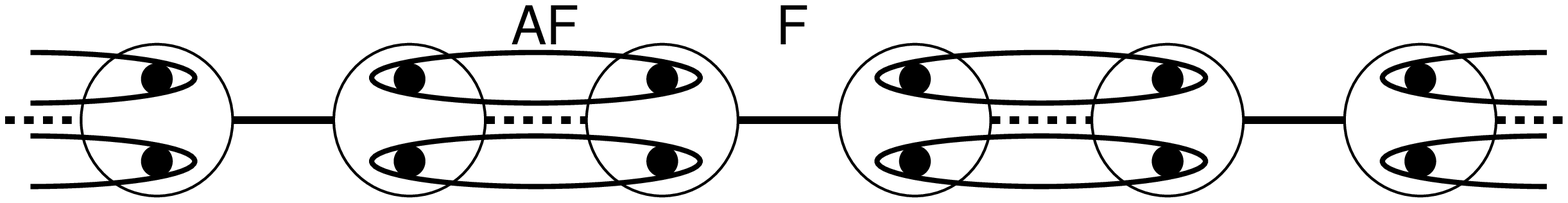}
   \end{center}
   \begin{center}
      \raisebox{0.2cm}{\texttt (c) }\includegraphics[scale=0.25]{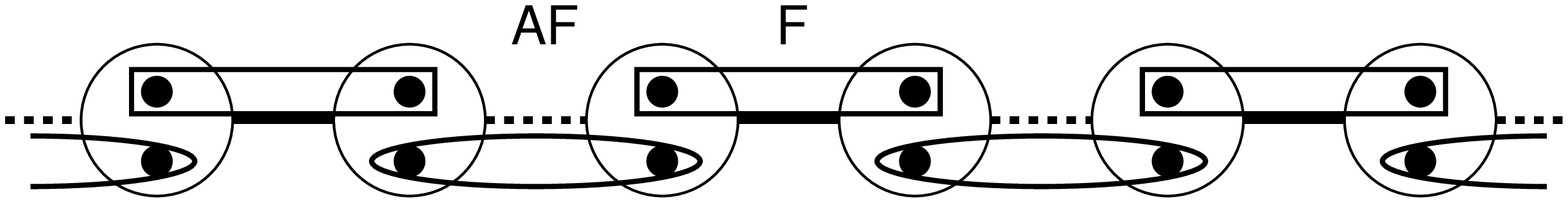}~~~~~
      \raisebox{0.2cm}{\texttt (d)}\includegraphics[scale=0.25]{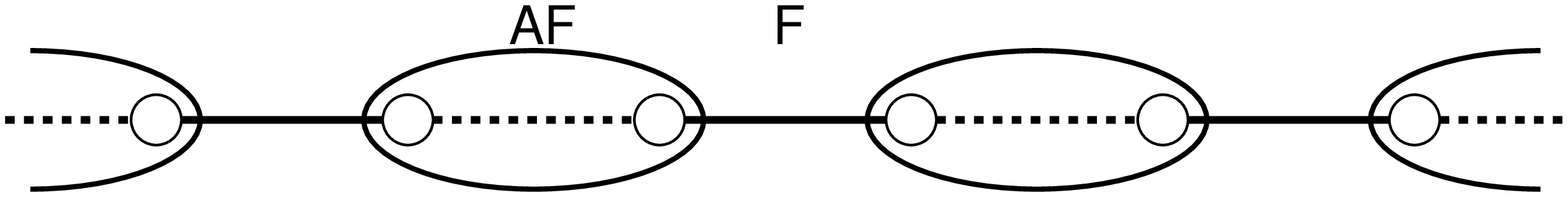}
   \end{center}
   \begin{center}
      \raisebox{0.6cm}{\texttt (e)}\includegraphics[scale=0.25]{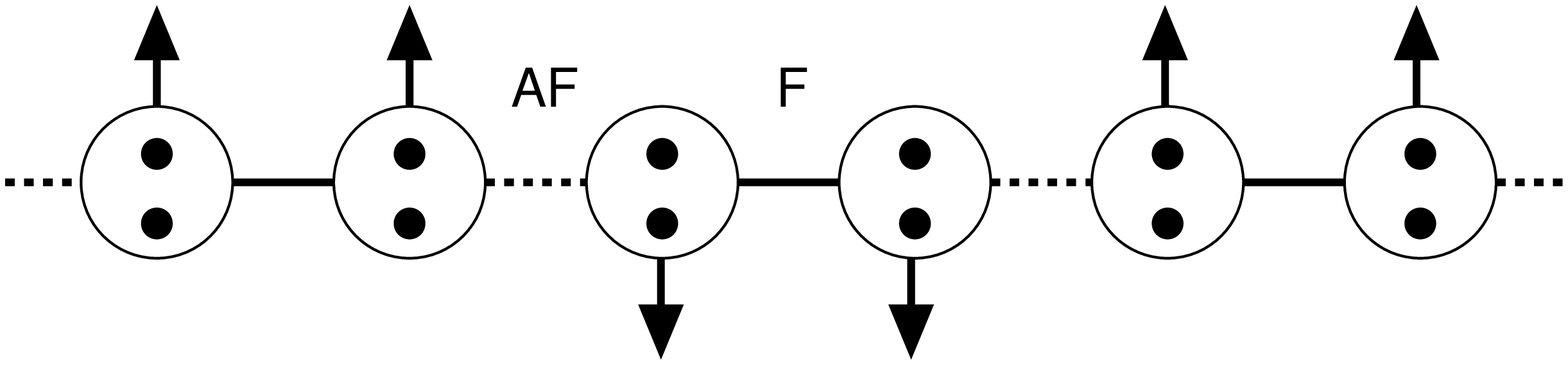}
   \end{center}
   \caption{Physical pictures of the (a) LD, (b) H, (c) ID, (d) SD and (e) N\'eel states.
            Big circles denote $S=1$ spins, while small dots $S=1/2$ spins.
            Ellipses represent singlet pairs, whilst rectangles ferromagnetically coupled spin pairs or clusters.
            The namings of these states are based on the pictures of the $T=2$ spin system except for the SD state.}
   \label{fig:fig3}
\end{figure}

Figure \ref{fig:fig2} shows our  GSPD on the
$\JAF$ versus $\DAF$ plane, 
which has been determined by using a variety of numerical methods based on the exact diagonalization
calculation data.  
This GSPD consists of six phases, which are
the $XY$1, LD, ID, H, spin-1 singlet dimer
(SD), and N{\'e}el (N) phases.  
The physical pictures of latter five states are sketched in Fig.\ref{fig:fig3}.
Among them, the LD, H, and SD phases are the trivial phases, while the ID phase is the SPT phase. 
Interestingly, the former three are smoothly connected without any quantum phase transitions
between the LD and H phases and between the H and SD phases, and therefore
they belong to the same phase.  
It is also emphasized that the ID phase appears in a wider region compared with the case of the GSPD
of the Hamiltonian (1) \cite{tone,oka1,oka2}.

We now explain how to determine numerically the phase boundary lines in
the GSPD shown in Fig.\ref{fig:fig2}.
We denote, respectively,
by $E_0^{\rm P}(L,M)$ and $E_1^{\rm P}(L,M)$, the lowest and second-lowest
energy eigenvalues of the Hamiltonian~${\cal H}$ under the periodic boundary
condition within the subspace characterized by $N$ and $M$, 
where $N(=4$, $8$, $12$, $16)$ is the total number of spins in the system and
$M(=0$, $\pm 1$, $\cdots$, $\pm N)$ is the total magnetization.
We also
denote by $E_0^{\rm T}(L,M,P)$ the lowest energy eigenvalue of $\cH$ under
the twisted boundary condition within the subspace characterized by $N$, $M$,
and $P$, where $P(=+1$, $-1)$ is the eigenvalue of the space inversion
operator with respect to the twisted bond. 
We numerically calculate
these energies by means of the exact diagonalization method.
In the following way, we evaluate the finite-size critical values of $\JAF$
(or $\DAF$) for various values of $\DAF$ (or $\JAF$)
for each phase transition.  Then, the phase boundary line for the transition
is obtained by connecting the results for the $N \to \infty$
extrapolation of the finite-size critical values.

Firstly, the phase transition between the LD and ID phases and that between
the ID and H phases are of the Gaussian type.  Therefore, as is well known,
the phase boundary lines can be accurately estimated by Kitazawa's level
spectroscopy (LS) method \cite{kitazawa-ls}. 
Namely, we numerically solve the
equation,
\begin{equation}
    E_0^{\rm T}(N,0,+1) = E_0^{\rm T}(N,0,-1)
\end{equation}
to calculate the finite-size critical values.  It is noted that, at the
\hbox{$N\!\to\!\infty$} limit,
$E_0^{\rm T}(N,0,+1)\!>\!E_0^{\rm T}(N,0,-1)$ in the ID phase
and $E_0^{\rm T}(N,0,+1)\!<\!E_0^{\rm T}(N,0,-1)$ in the LD and H
phases.

Secondly, the phase transitions between one of the LD, ID, H, and SD phases
and the $XY$1 phase are of the Berezinskii-Kosterlitz-Thouless \cite{berezinskii,kt,kosterlitz} type. 
Then,
the phase boundary line can be accurately estimated by the LS method developed by
Nomura and Kitazawa \cite{nomura-kitazawa-ls}.
Then, we solve the following equation to calculate the finite-size critical values:
\begin{equation}
    E_0^{\rm P}(N,2) = E_0^{\rm T}(N,0,P)\,,
\end{equation}
where \hbox{$P\!=\!-1$} or \hbox{$P\!=\!+1$} depending upon whether the
transitions are associated with the ID phase or with the LD, H, and SD
phases.

Lastly, since the phase transitions between one of the LD, H, SD phases and
the N phase are the 2D Ising-type transition, the phase boundary line between
these two phases can be estimated by the phenomenological renormalization
group method \cite{nightingale}.  
Then, the finite-size critical values for this transition
are calculated by solving the equation,
\begin{equation}
    N\,\Delta_{00}^{\rm P}(L,0) = (N+4)\,\Delta_{00}^{\rm P}(L+4,0)\,,
\end{equation}
where 
$\Delta_{00}^{\rm P}(L,0) = E_1^{\rm P}(L,0) - E_0^{\rm P}(L,0)$.

\section{Perturbation Theory from the Strong Ferromagnetic Coupling Limit}

In the strong ferromagnetic coupling limit,
it is thought that the present system can be mapped onto the spin-2 model.
Here we take the unperturbed Hamiltonian as
\begin{eqnarray}
  &&{\cal H}^{(0)} = \sum_j  h^{\rm F(0)}_{2j-1,2j}\,, \\
  && h^{\rm F(0)}_{2j,2j+1} = -J_{\rm F} (S_{2j}^x S_{2j+1}^x + S_{2j}^y S_{2j+1}^y 
                       + S_{2j}^z S_{2j+1}^z)\,.
\end{eqnarray}
The ground states of $h^{\rm F(0)}_{2j,2j+1}$ are five-fold degenerate,
which are interpreted as isolated spin-2 states,
expressed by the spin-2 operator $\vT$.
We note that, 
if we include the $XXZ$ anisotropy in ${\cal H}^{(0)}$,
we cannot treat lowest five states as isolated spin-2 states,
In the lowest order perturbation theory,
we obtain
\begin{eqnarray}
    &&\cHeff^{T=2}
      = {\JAF \over 4}
      \left\{
        \sum_j \left(T_{j}^x T_{j+1}^x + T_{j}^y T_{j+1}^y + \DAF T_{j}^z T_{j+1}^z \right)
        + \sum_j \left( \tilde D_2 (T_j^z)^2 + \tilde D_4 (T_j^z)^4\right) 
      \right\}\,,  
    \label{eq:heff-T}    \\
    &&\tilde D_2 = {14D_2 - 4\beta \over 3\JAF} \,,\\
    &&\tilde D_4 = -{2D_2 \over 3\JAF} \,,
    \label{eq:effectiv-H}
\end{eqnarray}
where
\begin{equation}
    \beta \equiv \JF(\DF-1).
\end{equation}
It is interesting that $\tilde D_4 (T_j^z)^4$ term appears.
Since we have set $\DF=0.8$, and $D_2=-1/30$, 
it holds
\begin{equation}
   \tilde D_2 = {1 \over 9\JAF} \,,~~~~~
   \tilde D_4 = {1 \over 45\JAF}\,.
   \label{eq:tilde-D}
\end{equation}

\begin{figure}[ht]
   \begin{center}
      \includegraphics[scale=0.35]{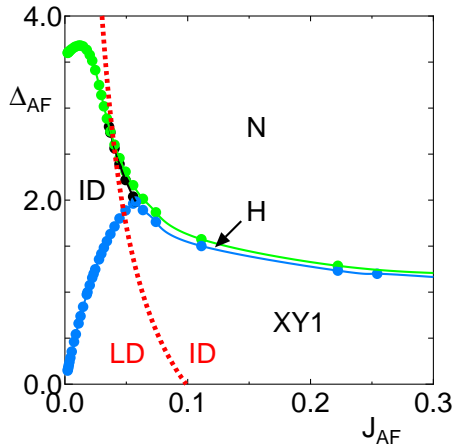}
   \end{center}
   \caption{The GSPD obtained by the perturbation theory in case of $\DF=0.8$, and $D_2=-1/15$.
            The red dotted line is the result by mapping onto the $\sigma = 1$ model,
            while other lines (black, blue and green) are that by mapping onto the $T = 2$ model.
            If we combine these results around $\JAF = 0.05$,
            the numerically obtained GSPD (Fig.\ref{fig:fig2}) is qualitatively explained.
           }
   \label{fig:perturb}
\end{figure}

Unfortunately,
the GSPD of the Hamiltonian (\ref{eq:heff-T}) with the parameter set given by eq.(\ref{eq:tilde-D})
has not been reported in the literature.
However, that with $\DF=0.8$, and $D_2=-1/15$ (namely, $-\tilde D_2 = \tilde D_4 = 1/60\JAF$)
is shown in Fig.3(a) of our previous paper\cite{oka3},
Thus, as the second-best plan,
we are going to compare our present GSPD with that in ref.\cite{oka3}.
The GSPD of Fig.3(a) of \cite{oka3} can be recasted into Fig.\ref{fig:perturb}.

When the $XXZ$ anisotropy of the ferromagnetic interaction is introduced (namely, when $\beta \ne 0$),
the five-fold degenerate states of ground states of $h^{\rm F(0)}_{2j,2j+1}$
are split into three levels with $T^z =0$, $T^z = \pm 1$ and $T^z = \pm 2$.
We note that this effect is expressed as the $\tilde D_2 (T_j^z)^2$ and $\tilde D_4(T_j^z)^4$ terms
in the effective Hamiltonian (\ref{eq:effectiv-H}).
For our parameter set ($\beta = -0.2$ and $D_2=-1/30$),
these energies are
\begin{equation}
    E_{\rm on-site}
    \equiv \JAF[\tilde D_2 (T^z)^2 + \tilde D_4(T^z)^4]
    =
    \begin{cases}
      0 \,,                                                             &(T^z=0) \,, \\
      \JAF\dstyle{\tilde D_4 \left( {\tilde D_2 \over \tilde D_4} + 1 \right)} = {2 \over 15}\,,   &(T^z = \pm 1) \,, \\
      \JAF\dstyle{\tilde D_4 \left( 4{\tilde D_2 \over \tilde D_4} + 16 \right)} = {4 \over 5}\,,  &(T^z = \pm 2) \,.
    \end{cases} 
\end{equation}
When $\JAF$ is much smaller than the splitting energy (for instance, $0.1$),
the states with $T^z=\pm 2$ will be strongly suppressed.
In this case,
it is appropriate to neglect the $T^z=\pm 2$,
which leads to the mapping onto the $\sigma=1$ spin system.
A straightforward calculation leads to
\begin{equation}
    \cHeff^{\sigma=1}
    = {3 \over 4\JAF}\left\{ \sum_j \left( \sigma_j^x \sigma_{j+1}^x + \sigma_j^y \sigma_{j+1}^y
       + {\DAF \over 3}\sigma_j^z \sigma_{j+1}^z\right)
      + {2 \over 45\JAF} \sum_j (\sigma^z)^2  \right\}.
    \label{eq:heff-sigma}
\end{equation}
The GSPD for the effective Hamiltonian (\ref{eq:heff-sigma})
was obtained by Chen et al.\cite{chen}.
The LD (trivial) and Haldane (SPT) states of the GSPD of Chen et al.
correspond to the LD (trivial) and ID (SPT) states of the GSPD of the present model, respectively.
By recasting the GSPD of Chen et al. leads to the red dotted line in Fig.\ref{fig:perturb}.
Thus, Fig.\ref{fig:perturb} quantitatively explains our numerical result shown in Fig.\ref{fig:fig2}.

\section{Concluding Remarks}

We have investigated the GSPD of an $S=1$
ferromagnetic-antiferromagnetic bond-alternating chain with the $XXZ$ and the on-site anisotropies
by using mainly numerical methods.
In the GSPD,
there appear the $XY$1, the large-$D$ (LD), the intermediate-$D$ (ID), 
the Haldane (H), the spin-1 singlet dimer (SD), and the N\'eel phases.  
Among them, the LD, the H, and the SD phases are
the trivial phases, while the ID phase is the SPT phase. 
We have also developed the perturbation theory from the strong ferromagnetic coupling limit
to map onto the $T=2$ effective model,
which quantitatively explains the numerically obtained GSPD.

We can see a considerably wider region of the ID phase in Fig.\ref{fig:fig2}
than that for the model described by the Hamiltonian (\ref{eq:Ham-1}),
in which the ID phase was numerically observed for the first time.
The reason for the wider ID region in Fig.\ref{fig:fig2} is the existence of the
$\tilde D_4 (T_j^z)^4$ term in eq.(\ref{eq:heff-T}).
We have already shown that the addition of the $D_4 (T_j^z)^4$ term with $D_4 >0$
to the Hamiltonian (\ref{eq:Ham-1}) drastically widen the ID region \cite{oka3}.
We believe that the finding of the wider ID region in our GSPD provides a guiding principle to
find or synthesize real materials in which the ID phase could be experimentally observed. 

\section*{Acknowledgments}

This work was partly supported by JSPS KAKENHI, Grant Numbers 16K05419, 
16H01080 (J-Physics) and 18H04330 (J-Physics). 
A part of the computations was performed using 
facilities of the Supercomputer Center, 
Institute for Solid State Physics, University of Tokyo, 
and the Computer Room, Yukawa Institute for Theoretical Physics, 
Kyoto University.

\end{document}